\documentclass[prl,twocolumn,showpacs,amsmath,amssymb]{revtex4-1}

\usepackage{graphicx}% Include figure files
\usepackage{dcolumn}% Align table columns on decimal point
\usepackage{bm}% bold math
\def\3he{$^3$He}
\def\4he{$^4$He}

\begin{document}

%\preprint{APS/123-QED}

\title{Mass flux and solid growth in solid $^4$He: 60 mK - 700 mK}% Force line breaks with \\

\author{M.W. Ray}
\author{R.B. Hallock}%
\affiliation{%
Laboratory for Low Temperature Physics, Department of Physics,\\
University of Massachusetts, Amherst, MA 01003
}%

\date{July 9, 2010}% It is always \today, today,

\begin{abstract}

 We use the thermo-mechanical effect to create a chemical potential difference between two liquid reservoirs connected to each other through Vycor rods in series with solid hcp \4he to confirm that a DC flux of atoms takes place below $\sim$ 600 mK, but find that the flux falls abruptly in the vicinity of 80 mK.  It is impossible to add density to a solid freshly made at 60 mK and samples freshly made at 60 mK do not allow mass flux, even when raised in temperature to 200 mK. Solids created above $\sim$ 300 mK and cooled to 60 mK accept added density and demonstrate finite mass flux.

\end{abstract}

\maketitle

Kim and Chan\cite{Kim2004a,Kim2004b} observed a shift in the resonant period of a torsional oscillator filled with solid \4he when cooled below $\sim 200$ mK. This shift was interpreted as due to mass decoupling and it was suggested that this was likely evidence for a supersolid phase\cite{Penrose1956,Andreev1969,Chester1970,Leggett1970} of solid \4he.  This interpretation has spawned considerable activity and debate.  It has become apparent that supersolidity should not be present in a perfect \4he crystal\cite{Ceperley2004,Clark2006,Boninsegni2006}; any possible superfluidity likely involves disorder. Recent work questions the supersolid interpretation and implies that many experiments carried out to date may, in fact, show no direct evidence for superfluid behavior\cite{Syshchenko2010,Reppy2010}.

If a supersolid does exist, such a solid might be expected to allow a mass flux through it. Attempts to create such flow in solid \4he in confined geometries by directly squeezing the solid lattice have not been successful\cite{Greywall1977,Day2005,Day2006,Rittner2009}.  We took a different approach and by injecting atoms from the superfluid have demonstrated mass transport through a cell filled with solid\cite{Ray2008a,Ray2009b,Ray2010a} at temperatures below $\sim$ 600 mK. We have also demonstrated an ability to add density to a solid\cite{Ray2010a}. More recently we have used the thermo-mechanical effect to create a chemical potential difference between reservoirs in series with solid helium and observe a rate-limited change of the thermo-mechanical pressure, which confirms the presence of a flux of atoms through the solid-filled cell\cite{Ray2010b}.

In the present work we use the thermo-mechanical effect to document this flux to much lower temperatures and find that a strong temperature dependence is present below $\sim$ 80 mK.  We also observe a strong temperature dependence to our ability to change the solid density (the so-called ``isochoric compressibility"\cite{Soyler2009}, $\chi$), in the same temperature range, with $\chi$ apparently zero at 60 mK for some samples, as demonstrated by an inability to move a sample freshly-made at this temperature off the melting curve by injection of atoms.

\begin{figure}[b]
\resizebox{2 in}{!}{
\includegraphics{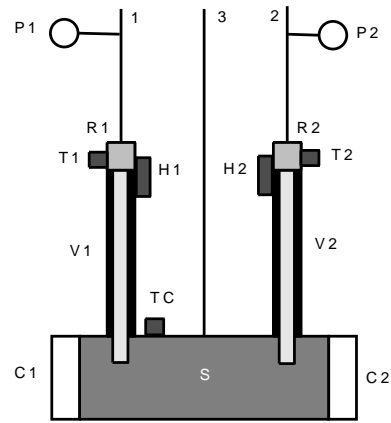}}
\caption{\label{fig:cell} Schematic diagram of the apparatus. The pressure of the solid is measured by capacitance strain gages \cite{Straty1969} C1 and C2, the solid temperature is measured by thermometer TC and the pressures of the reservoirs are measured by pressure gauges P1 and P2 located outside the cryostat.}
\end{figure}

Figure \ref{fig:cell} shows the apparatus used for this work\cite{Ray2008a,Ray2009b,Ray2010a}.  With solid \4he in region S and the pressure $\lesssim$ 37 bar, there is superfluid liquid helium in the Vycor rods\cite{Beamish1983,Lie-zhao1986,Adams1987}, V1, V2.  A temperature gradient is present across the Vycor; the cell remains at a low temperature, measured by thermometer TC.   A chemical potential difference, $\Delta \mu$, can be imposed between the Vycor rods by the creation of a pressure difference $P2 - P1 \neq$ 0 by adding or subtracting atoms from the superfluid-filled reservoirs R1 or R2 using lines 1 or 2.  A chemical potential difference can also be imposed by use of  heaters H1 or H2, thus creating a temperature difference, $\Delta T = T1 - T2$, between the two reservoirs.  The resulting change in the fountain pressure\cite{Ray2010b} between the two reservoirs, $\Delta P_f = \int_{T1}^{T2}\rho S dT$, results in a mass flux through the solid-filled cell to restore equilibrium. Here $\rho$ is the density and $S$ is the entropy. This alternate approach is advantageous since the number of atoms in the apparatus is fixed, thus allowing repetitive measurements on a sample without the provocation of net added mass, which would change the pressure of the solid.

 To fill the cell initially, the helium (commercial grade $\sim$300 ppb \3he) is introduced through line 3 (Figure \ref{fig:cell}) and does not pass through the Vycor rods, thus preserving the $^3$He concentration in the sample.  To grow the solid at constant temperature, we begin with the pressure in region S just below the bulk melting curve for \4he and then add atoms through lines 1 and 2.  For example, at $TC$ = 350 mK we can continue to add atoms to the solid at pressures greater than the bulk freezing pressure of the helium, and grow a solid sample\cite{Ray2010a} to a selected pressure and then change $TC$.  At low temperatures we have found a change in the growth behavior and will return to this point later in this report.

With solid helium in region S, we close the three fill lines and use H1 or H2 to vary $T1$ or $T2$, create a chemical potential difference between the reservoirs and then measure the resulting changes\cite{Ray2010b} in the pressures $P1$ and $P2$.  We also monitor $\emph{in situ}$ pressure gauges C1 and C2.  We have observed that the pressure-temperature relationship given by the relation for $\Delta P_f$ noted above, is obeyed\cite{Ray2010b} by $\Delta P $ for the case of liquid or solid helium in the region S.

\begin{figure}
\resizebox{3.5 in}{!}{
\includegraphics{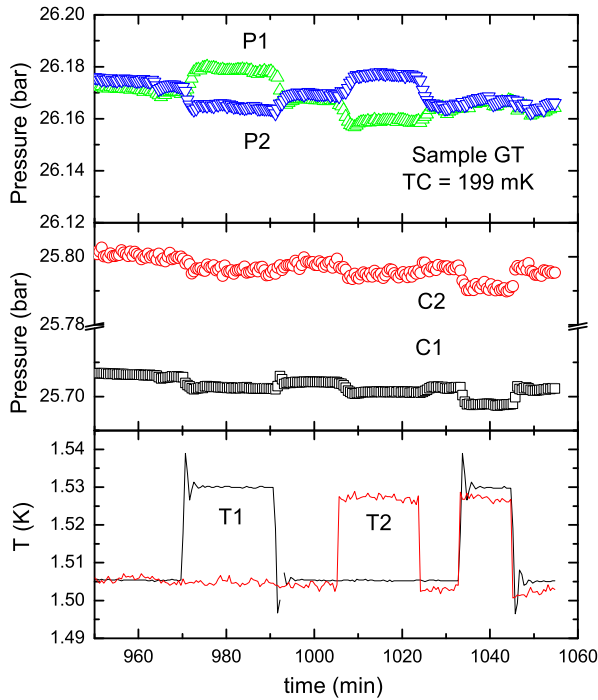}}
\caption{ \label{fig:rate1} (color online) Sample GT; grown at $TC$ = 300 mK, cooled to 60 mK and then warmed to $TC$ = 199 mK.  Use of heaters H1 or H2, results in a change in $T1$ or $T2$.  The changes in $P1$ and $P2$ are linear in time. Small decreases are also seen in $C1$ and $C2$.  Application of increases in temperature simultaneously to the reservoirs R1 and R2 results in a small change in $P1$ and $P2$, and a more substantial decrease in $C1$ and $C2$.   }
\end{figure}

Application of a sequence of temperature differences between the two reservoirs results in the creation of  fountain pressures and equilibrium is restored by the flow of atoms.   In Figure \ref{fig:rate1} we show the pressures, P1, P2, measured above each reservoir as a function of time that results from the imposition of temperature differences between reservoirs R1 and R2.  The reservoir pressures change in response to the applied temperature changes.  We believe that atoms move through the solid; for example, with $T2 > T1$, from reservoir 1 to 2.  For $\vert T1-T2 \vert \neq$ 0 in Figure \ref{fig:rate1}, C1 and C2 decrease by $\approx$ 2.5 mbar, which indicates that the solid is also a source of atoms.

When both heaters are simultaneously employed, the absence of a temperature difference between the reservoirs should result in the absence of a fountain pressure between the two.  But, with each of $T1$ and $T2$ increased, there will be an increase in the fountain pressure between the reservoirs and the solid in the sample cell.  Thus, one might expect atoms to move from the cell to the reservoirs to restore equilibrium.  An example of the behavior observed for such an experiment is shown in Figure \ref{fig:rate1}, where it is observed that there is only a small increase in the pressures $P1$ and $P2$ (Figure \ref{fig:rate1}, near t = 1040 min).  But, to restore equilibrium atoms appear to be taken from the solid as evidenced by the changes in the pressures seen on the in situ gauges C1 and C2 ; $C1$ and $C2$ each shift by $\approx$ 5 mbar.  Thus, the solid behaves as a reservoir and a source for atoms.

\begin{figure}
\resizebox{3.5 in}{!}{
\includegraphics{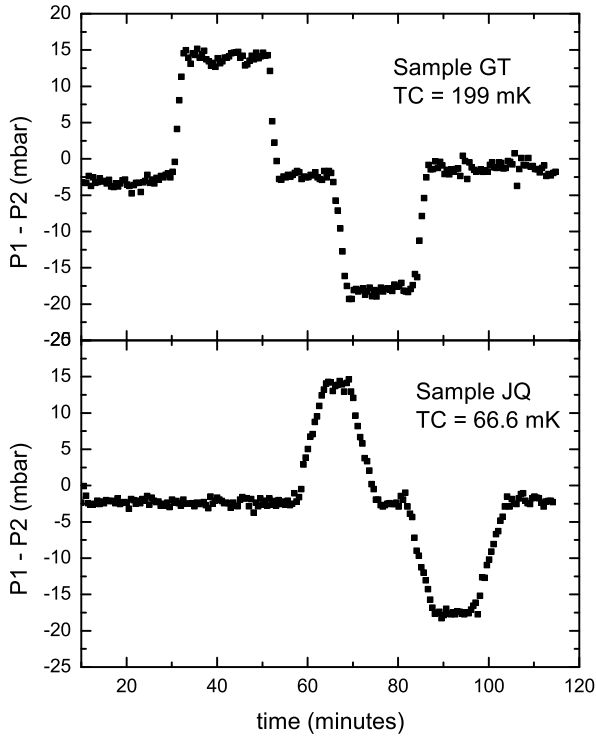}}
\caption{ \label{fig:rate2} (a) The differential pressure $P1 - P2$ for the data shown in Figure \ref{fig:rate1}; sample GT, $TC$ = 199 mK.  (b) Sample JQ, $TC$ = 66.6 mK; $P$ = 26.07 bar, grown at 316 mK.  The changes in $P1 - P2$ are linear in time, with $d(P1-P2)/dt$ (a measure of the flux) typically limited by the presence of the solid. The linear segment slopes for such data from various samples are later referred to as A, B, C and D, respectively, from left to right.}
\end{figure}

Figure \ref{fig:rate2} illustrates the differential pressure $\Delta P = P1-P2$ that results from the imposition of these temperature changes to the reservoirs.  The linear change in $P1-P2$ with time following application of $\vert T1-T2 \vert \neq$ 0 is evident and is a measure of the flux of atoms.  The size of this flux typically is constant as equilibrium is restored and in most cases is limited by the solid; thus there is a temperature-dependent critical flux.  A change in the apparatus has now allowed us to reach temperatures below 60 mK and thus study this flux as a function of temperature over a broad range.

\begin{figure}
\resizebox{3.5 in}{!}{
\includegraphics{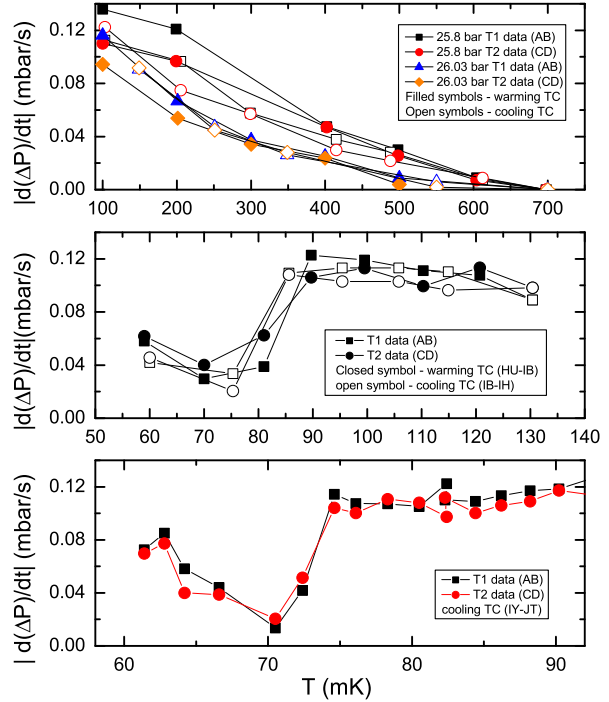}}
\caption{ \label{fig:slope3} (color online) (a) Rate of change of $\Delta P$ vs $TC$ for $TC$ $\geqslant$ 100 mK as determined from data like that shown in Figure \ref{fig:rate2}. (b) Rate of change of $P1-P2$ vs. $TC$ $\leqslant$ 130 mK  for both increasing and decreasing $TC$ (sequences HU-IB, IB-IH; ($C1+C2$)/2 = 26.06 bar).  (c) Finer grid of data taken on cooling for another sample (sequence IY - JT; ($C1+C2$)/2 = 26.01 bar).  Two data points in (c) following a \4he transfer have been removed.  In all cases shown here the absolute values of the slopes A(C) and B(D) (defined in Figure \ref{fig:rate2}, caption) have been averaged. Note that the midpoint of the steep portion of the data has shifted slightly for the two different samples.}
\end{figure}

As $TC$ is reduced, $d(\Delta P)$/dt vs $TC$, determined from such data, rises smoothly from zero near 600 mK,  Figure \ref{fig:slope3}a.  The critical flux through our superfluid-filled Vycor is measured to be no less than 0.14 mbar/sec.  Near $TC$ $\approx$ 75 mK such slopes fall precipitously and may develop structure and hysteresis. Results from additional experiments with smaller $TC$ temperature steps with different samples are shown in Figure \ref{fig:slope3}b,c.  In the temperature range where the temperature dependence is strong, measurements of the slope sometimes reveal time-dependent changes, which suggests that dynamic processes likely are present in the solid.  We attribute the temperature-dependent changes (and temporal changes) in the ability to carry the flux through the solid to changes in the solid itself, perhaps due to changes in structures within the solid.  We note that this characteristic temperature is similar to the temperature at which the dissipation peak appears in torsional oscillator experiments and where a change of the stiffness of the solid has been observed\cite{Day2007}.

The cause of this behavior of the mass flux through the solid-filled cell is still not known, but the present work suggests that there may be validity to some of the more recent suggestions.  One possibility is the presence of superfluid cores along edge dislocations that climb as proposed by Soyler {\it et al}\cite{Soyler2009}. This proposal explains two aspects of our experiments: (1) the flux of atoms though solid helium below an onset temperature ($\sim$ 600 mK) and (2) the growth of the density of the solid at constant volume in the presence of a chemical potential gradient, growth that is predicted to become impossible below a characteristic low temperature\cite{Aleinikava2008,Aleinikava2010}.

\begin{figure}
\resizebox{3.5 in}{!}{
\includegraphics{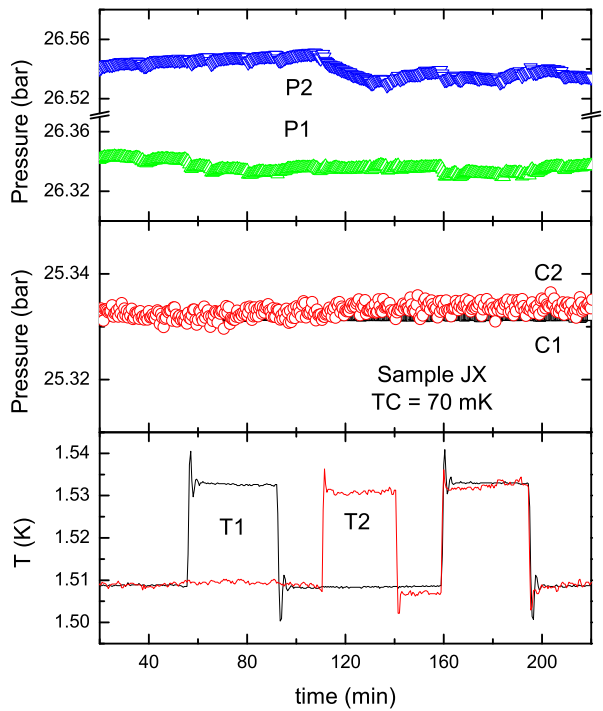}}
\caption{ \label{fig:graph5} (color online) Sample JX, created at $TC$ = 60 mK, on the melting curve, here with $TC$ = 70 mK.  Application of thermal energy to either H1 or H2 or both, results in a change in $T1$ or $T2$ or both.  There are no resulting significant changes in $P1, P2, C1$ or $C2$ that are reminiscent of those seen, for example, in Figure \ref{fig:rate1}.   A small DC drift correction has been applied to the $P2$ data.}
\end{figure}

Next we return to the growth behavior of the solid.  Typically with the sample cell filled with liquid at about 24 bar and $T$C above about 300 mK,  an increase in the pressure to lines 1 and 2 results in the growth of a solid on the melting curve followed by an increase in the density of the solid as it moves off the melting curve\cite{Ray2010a}. This is an example of what has been termed\cite{Soyler2009} a ``syringe experiment" and is the protocol we have often used to increase the pressure and density of the solid in our fixed volume. Thus, $\chi \neq$ 0 at temperatures where this behavior is observed.
 A solid created at, for example, 350 mK can be increased in density by injection of atoms simultaneously through lines 1 and 2 in a few tens of minutes (with, e.g., $\Delta C1, C2 \sim$ 0.5 bar).  However, no increase in the density of a solid freshly created at 60 mK is evident (after dwell times as long as $t_w$ $\approx$ 24 h); it will not grow above the melting curve, i.e., there is no syringe effect, $\chi$ = 0.  This remains true when such a sample is warmed to 70, 100, 200 mK ($t_w$ $\approx$ 16 h) and in a different sample 250 mK ($t_w$ $\approx$ 17 h); $\chi \neq$ 0 at $\approx$ 300 mK.  Thus, for such a sample $\chi$ is vanishingly small, but becomes finite after an increase in temperature above $\approx $ 300 mK. In addition, if fountain measurements of the sort shown in Figure \ref{fig:rate1} are attempted for such a freshly made cold sample, no changes in $P1, P2, C1$ or $C2$ reminescent of those seen in Figure \ref{fig:rate1} are present, Figure \ref{fig:graph5}; there is no evidence for flow. Thus, such samples do not demonstrate supersolid-like behavior. But, if such a sample is warmed to above 300 mK, $\chi >$ 0 and it can be grown to higher density. When returned to lower temperatures, it demonstrates flow behavior such as as shown in Figures \ref{fig:rate1} - \ref{fig:slope3}.  Thus, there is hysteresis.

 We believe that this work provides evidence that superfluid-like behavior is present in the cell filled with
solid helium, but the presence of such behavior at low temperature depends importantly on the method of preparation of the sample. Why this is so remains an open question.  This work suggests that a true understanding of these interesting phenomena in solid helium involves an interplay between the structural properties of the solid and the presence of superfluidity along the structures that carry the flux; structural properties alone are likely not adequate to explain the behavior we have observed.

In summary, we have used the fountain effect to apply chemical potential differences across samples of solid helium and seen clear evidence for substantial changes in the ability of solid \4he samples to carry a flux of helium atoms as a function of temperature. In particular, there is a substantial change in the flux at and below about 75 mK. And, at these lower temperatures $\chi$ for freshly made samples vanishes and so does the ability of such samples to allow mass flux.

 We thank A. Kuklov, J. Machta and B. Svistunov for helpful discussions.  This work was supported by NSF DMR  07-57701 and 08-55954 and by Research Trust Funds administered by the University.

\bibliography{ref}% Produces the bibliography via BibTeX.

%Included for Gather Purpose only:
%input "C:\Documents and Settings\mike\localtexmf\bibtex\bib\ref.bib"
\end{document}